\documentclass[aps, prd, showpacs, superscriptaddress, ctexart, nofootinbib, twocolumn]{revtex4}

\usepackage{amssymb, amsmath, bm, dcolumn, epsf, graphicx, latexsym, slashed, simplewick}

\usepackage{color}

\def\be{\begin{equation}}
\def\ee{\end{equation}}
\def\bea{\begin{eqnarray}}
\def\eea{\end{eqnarray}}
\begin{document}

\title{Cosmology of the Spinor Emergent Universe and Scale-invariant Perturbations}

\author{Yi-Fu Cai\footnote{Email: yifucai@physics.mcgill.ca}}
\affiliation{Department of Physics, McGill University, Montr\'eal, QC, H3A 2T8, Canada}

\author{Youping Wan\footnote{Email: wanyp@ihep.ac.cn}}
\affiliation{Institute of High Energy Physics, Chinese Academy of Sciences, Beijing 100049, China}

\author{Xinmin Zhang\footnote{Email: xmzhang@ihep.ac.cn}}
\affiliation{Institute of High Energy Physics, Chinese Academy of Sciences, Beijing 100049, China}

\pacs{98.80.Cq}

\begin{abstract}
A nonsingular emergent universe cosmology can be realized by a nonconventional spinor field as first developed in \cite{Cai:2012yf}. We study the mechanisms of generating scale-invariant primordial power spectrum of curvature perturbation in the frame of spinor emergent universe cosmology. Particularly, we introduce a light scalar field of which the kinetic term couples to the bilinear of the spinor field. This kinetic coupling can give rise to an effective ``Hubble radius" for primordial fluctuations from the scalar field to squeeze at large length scales as well as to form a nearly scale-invariant power spectrum. We study the stability of the backreaction and constrain the forms of the coupling terms. These almost scale-independent fluctuations are able to be transferred into curvature perturbation after the epoch of emergent universe through a generalized curvaton mechanism and thus can explain cosmological observations.
\end{abstract}

\maketitle

\section{Introduction}

Inflation has become the most prevailing and successful model of describing physics of very early universe \cite{Guth:1980zm} (also see \cite{Starobinsky:1980te} for early works). It provides instructive clues of explaining conceptual issues of the hot Big Bang cosmology. In particular, inflationary cosmology has predicted a nearly scale-invariant power spectrum of primordial curvature perturbation which was confirmed by a series of Cosmic Microwave Background (CMB) observations \cite{Ade:2013zuv}. However, it was pointed out in \cite{Borde:1993xh} that inflationary cosmology still suffers from the problem of the big bang singularity where conventional knowledge about mathematics and fundamental physics does not apply.

In past decades of years, alternative scenarios to inflationary cosmology remain to draw a lot of attention from cosmologists which can be the same successful as inflation in explaining very early universe and also avoids the initial big bang singularity (namely, see \cite{Brandenberger:2011gk} for a recent comprehensive review). These models can be divided into two categories. One is the bouncing cosmology in which the universe begins the evolution with a contracting phase and experiences a nonsingular bounce to connect the regular thermal expansion \cite{Mukhanov:1991zn, Cai:2012va, Cai:2013vm}. Amongst many bounce models, the matter bounce \cite{Wands:1998yp, Finelli:2001sr} and the Ekpyrotic cosmology \cite{Khoury:2001wf, Lehners:2008vx} are two representative scenarios as an explanation for the origin of the CMB and the Large Scale Sturcutre (LSS) of the universe.

In the literature there are various proposals of obtaining nonsingular bounces, namely, by modifying the gravitation theory as in Horava-Lifshitz gravity \cite{Calcagni:2009ar, Brandenberger:2009yt} and in torsion gravity \cite{Cai:2011tc}, by introducing certain Null Energy Condition (NEC) violating fields such as ghost condensate \cite{Buchbinder:2007ad, Creminelli:2007aq} or Galilean matter \cite{Qiu:2011cy}. A bouncing phase can originate from the quantum spacetime structure such as in loop quantum cosmology \cite{Ashtekar:2011ni}. Particularly, in order to realize a nonsingular bounce within the frame of Einstein gravity, it was shown in \cite{Cai:2007qw} that the equation-of-state (EoS) of the universe has to effectively cross the cosmological constant boundary for a while at early times, which is the so-called Quintom scenario \cite{Feng:2004ad}. This type of bounce models was studied in detail in \cite{Cai:2007zv} and later was reviewed in \cite{Cai:2009zp}. The combination of matter bounce and Quintom scenario was achieved by virtue of a Lee-Wick scalar in \cite{Cai:2008qw}. This model nicely demonstrates that cosmological perturbations generated in contracting phase can evolve through the bouncing phase smoothly and eventually give rise to a scale-invariant power spectrum of observable interest. The perturbation theory of bouncing cosmology has recently been greatly developed in a series of works, including the investigation of primordial non-gaussianities \cite{Cai:2009fn}, the study of entropy fluctuations \cite{Cai:2011zx}, and related reheating period \cite{Cai:2011ci}. The dynamics of cosmological perturbations within the pure Ekpyrotic cosmology were extensively studied in \cite{Xue:2010ux, Xue:2013bva, Koehn:2013upa}.

A second interesting nonsingular paradigm of very early universe is the so-called emergent universe cosmology \cite{Brandenberger:1988aj, Ellis:2002we}, in which our universe was emergent from a non-zero minimal length scale and experienced a sufficiently long period of quasi-Minkowski expansion and then began the normal big bang expansion. Motivated by string theory, the scenario of emergent universe can be achieved in the string gas cosmology due to the Hagedorn phase of a thermal system composed of a number of fundamental strings \cite{Brandenberger:1988aj}. Recently, the proposal of the Galilean model gives rise to the Galilean Genesis which also leads to the emergent universe scenario with an end of a big rip \cite{Creminelli:2010ba}. Phenomenologically, the study of the causal generation of primordial perturbation was developed in the conformal cosmology \cite{Rubakov:2009np}, the pseudo-conformal cosmology \cite{Libanov:2011zy, Hinterbichler:2011qk}, and later the cosmology of Galilean Genesis \cite{LevasseurPerreault:2011mw}, respectively. Moreover, various deformed versions of the emergent universe cosmology were analyzed in the literature, such as the processes of slow contraction \cite{Khoury:2001wf, Khoury:2009my} and slow expansion \cite{Piao:2003ty}, or connecting to inflationary cosmology \cite{Lehners:2012wz} as well as the braneworld scenario \cite{Zhang:2013ykz}.

The emergent universe cosmology, as it requires the Hubble rate approaching to zero in a infinitely long period in the past, implies that the NEC violating field is needed at very early times. This profound property was for the first time pointed out explicitly in \cite{Cai:2012yf}. It was verified by considering a parameterized Quintom fluid and then explicitly realized by introducing a cosmic spinor field. Embedding a nonconventional spinor field consistently into a curved spacetime such as our universe requires a very nice mathematical structure, and based on this advantage, one can reconstruct the potential of the cosmic spinor field according to the expected background evolution. This remarkable feature was earlier applied in the study of dark energy models \cite{Cai:2008gk} and in inflationary cosmology \cite{Feng:2012jm}. The analysis of \cite{Cai:2012yf} shows that an enough long period of emergent universe can be achieved if the potential of the spinor field has another minimum in the ultraviolet (UV) regime. Along with an extremely slow-expanding process during the quasi-Minkowski epoch, the spinor field would exit the state of tachyonic condensate and recovers the regular form of a massive fermion. Then the universe gracefully exits to the normal thermal expansion. This model, however, does not explain the origin of the CMB and LSS as observed in experiments since the perturbations of the spinor field form a blue spectrum.

In this paper we study the possibility of generating a nearly scale-invariant primordial power spectrum in the model of spinor emergent universe. We propose a generalized curvaton mechanism by introducing a second curvaton field which kinetically couples to the spinor field. Although during the emergent universe phase the background spacetime is almost static, the kinetic coupling term can change the friction term of the curvaton. Depending on the detailed forms of the kinetic coupling, the curvaton field could feel it is in a ``de-Sitter"-like background or a ``matter-contraction"-like one. In both two situations, a nearly scale-invariant power spectrum of iso-curvature perturbations can form. Afterwards, these iso-curvature fluctuations will be converted into curvature perturbation through the standard curvaton mechanism by assuming a process of adiabatic curvaton decay.

The paper is organized as follows. In Section \ref{Sec:model}, we briefly review the model of spinor emergent universe. Then, in Section \ref{Sec:chifield} we present two important issues existing in this model. To solve these issues, we introduce a curvaton scalar field kinetically coupled with the cosmic spinor field. Section \ref{Sec:pert} is devoted to the study of the primordial perturbations of this curvaton field. In particular, we perform a detailed analysis of the curvaton fluctuation and study the condition for producing a scale-invariant power spectrum. Then we reconstruct the form of the kinetic coupling as a function of the scalar bilinear of the spinor field with the stability issue being investigated. Numerical computation is performed in Section \ref{Sec:numerics} to examine the validity of the semi-analytical calculation in the end of this section. In Section \ref{Sec:curvaton} we study the conversion of the iso-curvature fluctuations into curvature perturbations by virtue of a generalized curvaton mechanism. We conclude with a discussion in Section \ref{Sec:conclusions}. Throughout the paper we take the sign of the metric to be $(+,-,-,-)$ and define the reduced Planck mass by $M_p = 1/\sqrt{8\pi G}$.

\section{The model of the spinor emergent universe}\label{Sec:model}

To start, we briefly review the emergent universe cosmology realized by a nonconventional spinor field minimally coupled with Einstein gravity \cite{Cai:2012yf}. The Dirac action in a curved spacetime background can be expressed as,
\begin{eqnarray}
 {\cal L}_\psi = e ~ [\frac{i}{2}(\bar\psi \Gamma^\mu D_\mu \psi - D_\mu \bar\psi \Gamma^\mu \psi) - U(\bar\psi\psi)] ~,
\end{eqnarray}
where $e$ is the determinant of the vierbein $e^a_\mu$. The Gamma matrices $\Gamma^\mu$ are defined under the Dirac-Pauli representation through $\Gamma^\mu \equiv e^\mu_a\gamma^a$, which satisfy the algebra $\{\Gamma^\mu, \Gamma^\nu\} = 2 g_{\mu\nu}$. Moreover, the covariant derivatives of the spinor field and its Dirac adjoint follow the relations below,
\begin{eqnarray}
 D_\mu\psi = \partial_\mu \psi + \Omega_\mu \psi ~, ~~D_\mu\bar\psi = \partial_\mu\bar\psi-\bar\psi\Omega_\mu ~,
\end{eqnarray}
where the spin connection $\Omega_\mu \equiv \frac{1}{2}e^\nu_a\nabla_\mu e_{\nu b}\Sigma^{ab}$ is defined. Additionally, we have introduced the generators of the spinor representation of the Lorentz group $\Sigma^{ab}=\frac{1}{4}[\gamma^a, \gamma^b]$.

By varying the Lagrangian with respect to the vierbein, we can derive the energy stress tensor as follows,
\begin{eqnarray}\label{Tmunu_spinor}
 T_{\mu\nu} = \frac{i}{2}[ \bar\psi\Gamma_{(\mu} D_{\nu)}\psi -D_{(\mu}\bar\psi\Gamma_{\nu)}\psi]
 - \frac{g_{\mu\nu}}{e} {\cal L}_\psi ~.
\end{eqnarray}
Further, one can vary the Lagrangian with respect to the spinor field and the adjoint, respectively, and then derive the equations of motion, which are expressed as
\begin{eqnarray}\label{eom_spinor}
 i\Gamma^\mu D_\mu\psi -U_{,\bar\psi\psi}\psi = 0~, ~~ i D_\mu\bar\psi\Gamma^\mu +U_{,\bar\psi\psi}\bar\psi = 0~,
\end{eqnarray}
where we have defined $U_{,\bar\psi\psi} \equiv \partial U/\partial(\bar\psi\psi)$. Note that, we assume the potential of the spinor field is only a function of the scalar bilinear $\bar\psi\psi$ in the case of interest.

Now we consider a spatially flat FRW universe with the metric of
\begin{eqnarray}
 ds^2 = dt^2 -a^2(t) d{\bf x}^2~,
\end{eqnarray}
and correspondingly, the vierbein are given by $e^\mu_0 = \delta^\mu_0$, $e^\mu_i = \frac{1}{a}\delta^\mu_i$. To assume the spinor field is only time-dependent, the equations of motion \eqref{eom_spinor} simply yield \cite{Cai:2012yf}
\begin{eqnarray}\label{bar_psi_psi}
 \bar\psi\psi = \frac{\cal N}{a^3}~,
\end{eqnarray}
within the FRW background with a positively defined constant ${\cal N}$. Moreover, the combination of the equations of motion \eqref{eom_spinor} and the energy stress tensor \eqref{Tmunu_spinor} determines the energy density and the pressure of the spinor field as follows,
\begin{eqnarray}
 \rho_\psi = U~,~~P_\psi = U_{,\bar\psi\psi} \bar\psi\psi - U~,
\end{eqnarray}
as well as the corresponding EoS
\begin{eqnarray}
 w_\psi \equiv \frac{P_\psi}{\rho_\psi} = -1+ \frac{U_{,\bar\psi\psi}\bar\psi\psi}{U}~.
\end{eqnarray}

The most profound property of this model is that, the EoS of the spinor field can cross $-1$ by changing the sign of $U_{,\bar\psi\psi}$ and thus naturally realizes the Quintom scenario. As has been pointed out in Ref. \cite{Cai:2012yf}, this property is the key element of achieving the emergent universe solution. We follow the specific solution obtained in \cite{Cai:2012yf} and briefly review the results.

The spinor emergent universe cosmology requires that the potential for the spinor field is a conventional one like $m\bar\psi\psi$ in the IR regime but becomes of a nontrivial form in the UV regime. Namely, if we consider in the UV regime, the potential is given by
\begin{eqnarray}\label{U_potential}
 U(\bar\psi\psi) \simeq \frac{3M_p^4}{{\cal C}^2}
 \left[ 1-a_{\rm i}^{\alpha {\cal C}} (\frac{\bar\psi\psi}{\cal N})^{\frac{\alpha {\cal C}}{3}} \right]^2~,
\end{eqnarray}
where $a_{\rm i}$ corresponds to the minimal value of the scale factor in the emergent universe cosmology. The coefficient ${\cal C}$ is significant in determining the energy scale of the occurrence of the emergent universe phase. The other coefficient $\alpha$, as will be discussed later, is an important parameter to characterize how fast the universe can exit the emergent universe phase.

In this model, the scale factor of the universe can be found to take the following asymptotical form
\begin{eqnarray}\label{scalefactor}
 a(t) \simeq a_{\rm E} \left( \frac{1+{\cal C} e^{\alpha Mp t}}{1+\frac{2}{3}\alpha {\cal C}} \right)^{\frac{1}{\alpha {\cal C}}} ~,
\end{eqnarray}
in the UV regime, but takes a regular power function of cosmic time after the emergent universe phase. The coefficient $a_{\rm E}$ is the value of the scale factor at the moment of exiting the emergent universe phase $t_{\rm E}$. It is related to the minimal value of the scale factor through $a_{\rm E}=a_{\rm i}(1+\frac{2}{3}\alpha {\cal C})^{\frac{1}{\alpha {\cal C}}}$. Note that, if the universe enters a matter dominated phase after the moment $t_{\rm E}$, then we have $a\sim (t-\tilde{t}_{\rm E})^{2/3}$. In this case, we can get the background EoS and the Hubble parameter as follows,
\begin{eqnarray}\label{Hubble_EU}
 w_\psi = -\frac{2\alpha}{3}e^{-\alpha M_p t}~,~~ H = \frac{M_p}{({\cal C}+\frac{3}{2}M_pt)+e^{-\alpha M_pt}}~.
\end{eqnarray}
From the expression of the Hubble parameter, one can learn that $H$ reaches the maximal value at the moment $t_{\rm E}$ with
\begin{eqnarray}\label{t_E}
 t_{\rm E} = \frac{\ln({2\alpha}/{3})}{\alpha M_p}~,~~H_{\rm E}=\frac{2\alpha M_p}{3+2\alpha {\cal C}+3 \ln\frac{2\alpha}{3}}~.
\end{eqnarray}
The above model can give rise to a emergent universe solution with a dust-like expansion following the emergent universe phase.

\section{Emergent universe involving a scalar field}\label{Sec:chifield}

The model of spinor emergent universe is analytically and numerical solvable on the background dynamics as illustrated in previous section. However, there exist two issues that deserve to be addressed. One is that the cosmological solution does not admit a reheating phase to drive the universe into thermal expansion. The other issue is related to the generation of primordial perturbations. In the second part of the appendix of \cite{Cai:2012yf}, it was found that during the emergent universe phase, the perturbations of the spinor field stay to be the vacuum fluctuations and cannot get squeezed since the background universe looks like a quasi-Minkowski one. Therefore, a pure spinor emergent universe scenario is not enough to explain the generation of the CMB and the LSS as observed.

To solve the above issues, we introduce a second scalar field $\chi$ which kinetically couples to the spinor field. The basic idea is similar to the curvaton mechanism that \cite{Lyth:2001nq}, the second field $\chi$ does not affect the background solution but is only responsible for the generation of primordial power spectrum in agreement with cosmological observations. In particular, we consider the Lagrangian of the curvaton field as follows,
\begin{eqnarray}
 {\cal L}_\chi = \frac{1}{2} {\cal Y}(\bar\psi\psi) \partial_\mu\chi\partial^\mu\chi - {\cal F}(\bar\psi\psi) V(\chi)~,
\end{eqnarray}
where ${\cal Y}$ and ${\cal F}$ are functions of the scalar bilinear $\bar\psi\psi$. The scalar field is free of ghost by requiring ${\cal Y}$ to be positive-definite.

Varying the Lagrangian with respect to the metric yield the effective energy density and the pressure of the curvaton field, which are given by,
\begin{align}\label{densitypressure}
 & \rho_\chi = \frac{{\cal Y}}{2} \dot\chi^2 + {\cal F}V(\chi)~,~ \nonumber\\
 & P_\chi = \frac{{\cal Y}}{2} \dot\chi^2 - {\cal F}V(\chi)
 -\frac{{\cal Y}_{,\bar\psi\psi}}{2} \bar\psi\psi\dot\chi^2 +{\cal F}_{,\bar\psi\psi}\bar\psi\psi V~,
\end{align}
where the dot represents for the derivative with respect to the cosmic time. The last two terms in the expression of the pressure are contributed from the variation with respect to the spinor field.

Additionally, we vary the Lagrangian with respect to $\chi$ and then derive the equation of motion for the curvaton field as follows,
\begin{eqnarray}\label{eom_chi}
 \ddot\chi + \frac{(a^3{\cal Y})^{\cdot}}{a^3{\cal Y}}\dot\chi +\frac{\cal F}{\cal Y}V_{,\chi} = 0~.
\end{eqnarray}
Note that, the friction term is not of the form $3H$ but also depends on the coefficient ${\cal Y}$ due to the kinetic coupling. This term could become important in the phase of emergent universe as the scale factor in nearly constant.

Since we expect this scalar plays the role of the curvaton that does not affect the background dynamics, the energy density of the $\chi$ field has to be subdominant in the emergent universe phase and so is the pressure of the curvaton. These impose theoretical constraints on the model under consideration.

\section{Cosmic perturbations}\label{Sec:pert}

\begin{figure}
\includegraphics[scale=0.3]{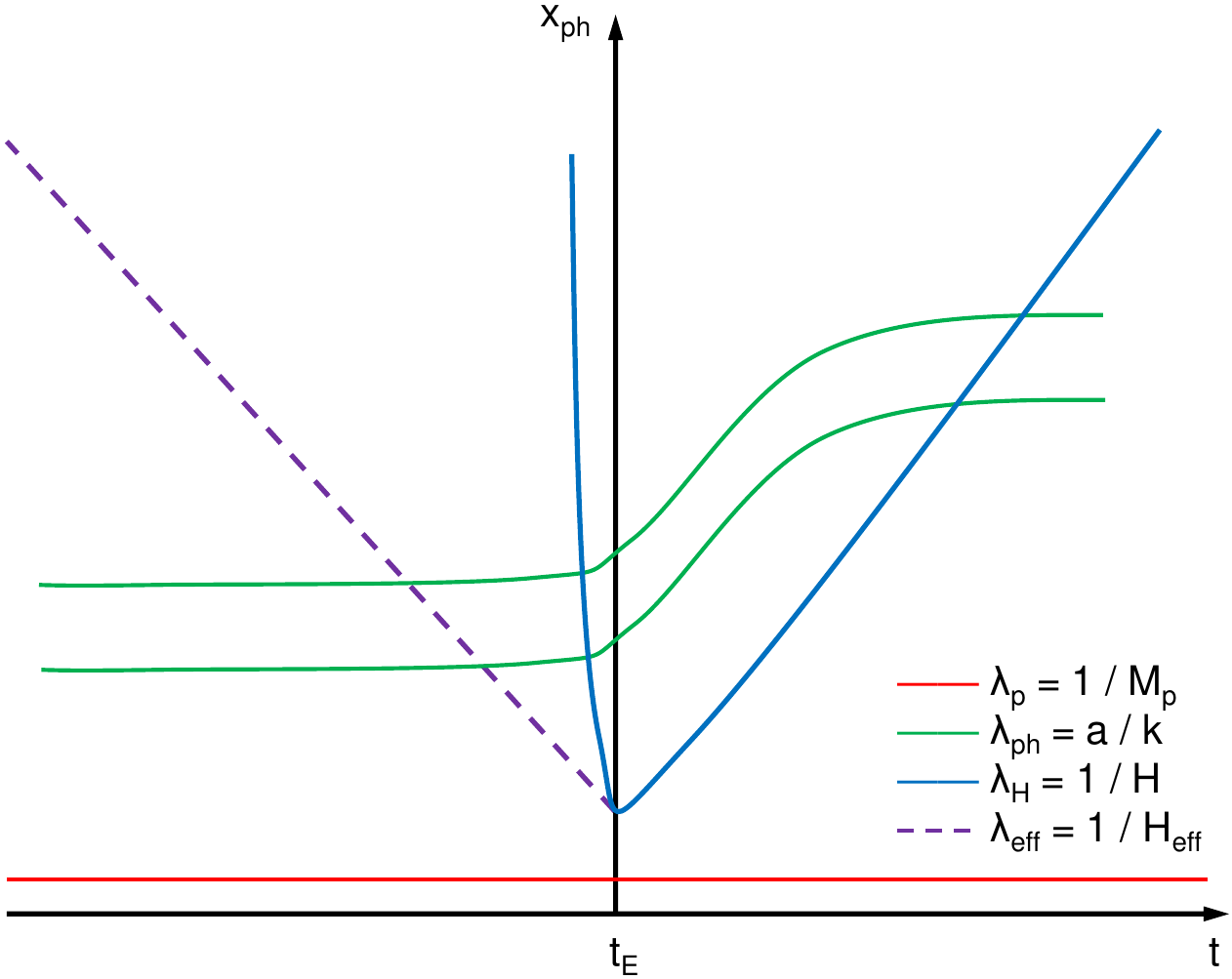}
\caption{A sketch plot of the dynamics of cosmological perturbations in the nonsingular emergent universe. The vertical axis is the physical spatial coordinate $x_{\rm ph}$, and the horizontal axis is the cosmic time $t$. The physical wavelength $\lambda_{\rm ph}=a/k$ of the mode with comoving wavenumber $k$ is depicted in green; the Hubble radius $\lambda_{H}=H^{-1}$ is depicted in blue; the effective ``Hubble radius" in the emergent universe phase is depicted by the purple dashed line; and the red line denotes the Planck length $\lambda_p=M_p^{-1}$. }
\label{Fig:sketchpert}
\end{figure}

In this section, we proceed to study the primordial perturbations generated from the curvaton field in the spinor emergent universe cosmology. At linear order, the Fourier mode of the field fluctuation evolves independently. Thus it is useful to track the evolution of each mode along with the background evolution. A causal generation mechanism of primordial fluctuations requires that, the physical wavelength of the fluctuation has to be sub-Hubble at very early times and then evolves to super-Hubble scale. In inflationary cosmology, it can be achieved by stretching the wavelength exponentially due to the inflationary process. In nonsingular bouncing cosmology, one can realize the similar scenario by decreasing the Hubble radius in contracting phase. We sketch the dynamics of primordial perturbations in the emergent universe cosmology in Fig. \ref{Fig:sketchpert}.

From Fig. \ref{Fig:sketchpert}, all the perturbation modes (the green lines) we are interested in originate in the sub-Hubble regime since the Hubble parameter converges to zero in the emergent universe phase. Along with the rapid growth of the Hubble parameter, the perturbations are able to exit the Hubble radius (the blue line) and propagation for a while and finally reenter at late times in the period of regular thermal expansion. The whole evolution of the fluctuations can be separated by the time moment $t_{\rm E}$ into two phases: the primordial ear of the emergent universe phase when the scale factor is nearly constant and the evolution in the post-emergent universe phase, respectively. However, one can easily observe that the wavelength of one perturbation mode does not change in the emergent universe epoch. It implies that the scale dependence of the power spectrum as expected cannot be explained by the mechanism of exiting the Hubble radius. Instead, we expect there exists an effective ``Hubble radius" (the purple dashed line) which can alternatively produce the scale dependence of the perturbation. As we will discuss later, this radius can be obtained by introducing the kinetic coupling on the curvaton field.

As a side remark, since the wavelength of the perturbation mode is frozen in the emergent universe phase, the minimal wavelength of observable interest could be comparable with the Hubble radius at the moment $t_{\rm E}$. Thus, it is larger than the Planck length (the red bottom line) provided that the energy scale of the emergent universe is sub-Planckian. This is a necessary requirement for the validity of the perturbation theory applied in the paper. This is a significant advantage compared to  inflationary cosmology where the wavelength of primordial fluctuation becomes smaller than the Planck length at the beginning if inflation lasted more than $70$ e-foldings \cite{Brandenberger:1999sw}. In our model, provided the energy scale of the universe at the bounce is lower than the Planck scale, then the physical wavelength of a perturbation mode corresponding to the current Hubble radius is in the far infrared, as shown in Fig. \ref{Fig:sketchpert}.

\subsection{Fluctuations of the curvaton field}

We perturb the curvaton field by $\delta\chi$ and work with the conformal time defined by $d\tau = dt/a$. It is important to focus on the canonically normalized perturbation variable:
\begin{eqnarray}
 v_\chi \equiv z \delta\chi~,~~z\equiv a{\cal Y}^{\frac{1}{2}}~.
\end{eqnarray}
in terms of which quantum vacuum initial conditions can be imposed at very early times. We work on the spatially-flat slice where the metric fluctuation of scalar type vanishes. On this slice, one can expand the action of the curvaton field up to quadratic order as follows,
\begin{eqnarray}
 S_\chi^{(2)} = \int d\tau d\mathbf{x}^3 \frac{1}{2} \left[v_\chi'^2-(\partial_i v_\chi)^2 - M^2 v_\chi^2 \right] ~,
\end{eqnarray}
with
\begin{eqnarray}
 M^2\equiv a^2\frac{\cal F}{\cal Y}V_{,\chi\chi} -\frac{z''}{z}~,
\end{eqnarray}
being an effective mass square term for the curvaton fluctuation.

To deal with the above perturbation system, it is useful to make a Fourier transformation and track the evolution of one Fourier mode. One can easily derive the equation of motion for the curvaton fluctuation with a fixed comoving wave number $k$ as,
\begin{eqnarray}\label{eom_vk}
 v_k''+(k^2+M^2)v_k = 0~.
\end{eqnarray}
To be explicit, we expand the effective mass square term as follows,
\begin{eqnarray}\label{M2_chi}
 M^2 = a^2\frac{\cal F}{\cal Y}V_{,\chi\chi} -\frac{a''}{a}-2{\cal H}\frac{({\cal Y}^{\frac{1}{2}})'}{{\cal Y}^{\frac{1}{2}}} -\frac{({\cal Y}^{\frac{1}{2}})''}{{\cal Y}^{\frac{1}{2}}}~,
\end{eqnarray}
where ${\cal H}\equiv a'/a$ is the conformal Hubble parameter. The last two terms are generated due to a kinetic coupling ${\cal Y}(\bar\psi\psi)$.
Note that, during the phase of emergent universe, the scale factor $a$ is almost constant and therefore the second and third terms in Eq. \eqref{M2_chi} become negligible. In order to make sure the model under consideration is free of tachyonic instability when siting on the vacuum state, one needs the first term $\frac{\cal F}{\cal Y}V_{,\chi\chi}$ to be positive-definite. As a result, the only possible term that can determine the squeezing process of the perturbation and the corresponding scale dependence comes from ${({\cal Y}^{\frac{1}{2}})''}/{{\cal Y}^{\frac{1}{2}}}$. Correspondingly, we would like to introduce an effective ``Hubble radius" as
\begin{eqnarray}
 \lambda_{\rm eff} \equiv a/ {\cal H}_{\rm eff} = a\frac{{\cal Y}^{\frac{1}{2}}}{({\cal Y}^{\frac{1}{2}})'} ~,
\end{eqnarray}
which can mimic the conventional Hubble radius in the causal mechanism of generating fluctuations in the primordial era.

Initially, the $k^2$ term dominates Eq. \eqref{eom_vk}. We can neglect the effective mass square term. Thus, the dynamics of the curvaton fluctuation exactly corresponds to a free scalar propagating in a flat spacetime. A natural choice of the initial condition takes the form of the Bunch-Davies vacuum:
\begin{eqnarray}\label{bunch_davies_vacuum}
 v_k \, \simeq \, \frac{e^{-ik\tau}}{\sqrt{2k}}~,
\end{eqnarray}
where we take $\tau\rightarrow -\infty$ as the initial moment. During the phase of emergent universe, we expect the quantum fluctuations can exit an effective ``Hubble radius" $\lambda_{\rm eff}$ and become classical perturbations. We specifically consider the case:
\begin{eqnarray}\label{V_chi}
 V(\chi)=\frac{1}{2}m^2\chi^2~,
\end{eqnarray}
with $m\ll H_{\rm E}$ to be required. We keep the form of ${\cal Y}$ to be undetermined. In order to obtain a nearly scale-invariant power spectrum, one expects that Eq. (\ref{eom_vk}) can be simplified as follows,
\begin{eqnarray} \label{squeeze}
 v_k'' + (k^2-\frac{2}{\tau^2})v_k \simeq 0~,
\end{eqnarray}
while the mass term has been neglected. The second term inside the parentheses on the left hand side of the above equation is the term which can lead to the squeezing of the curvaton fluctuations. The coefficient and time dependence is exactly that required to transform a blue vacuum spectrum into a scale-invariant spectrum.

To satisfy the above requirements, we find there exist two possible solutions to the kinetic coupling term ${\cal Y}$, which are given by,
\begin{eqnarray}\label{Y_sol}
  {\cal Y} = (-{\cal H}_{\rm E}\tau)^{-(2+2 \epsilon)} ~, ~~{\rm or}~~ {\cal Y} = (-{\cal H}_{\rm E}\tau)^{4+2\epsilon}~,
\end{eqnarray}
respectively. The coefficient $\epsilon$ is a small quantity which is introduced to slightly tune the scale dependence of the spectrum as will be discussed in the next subsection. In the following subsection, we study the generation of primordial power spectrum in both two cases.

\subsection{Scale-invariant spectrum}

Making use of the vacuum initial condition, we obtain an approximate solution to (\ref{squeeze}):
\begin{eqnarray} \label{sol}
 v_k \, \simeq \, \frac{e^{-ik\tau}}{\sqrt{2k}}(1-\frac{i}{k\tau})~,
\end{eqnarray}
where the effect brought by the $m^2$ term has been neglected. From this result, one can see the primordial curvaton fluctuations can be transformed from quantum vacuum fluctuations to classical perturbations without the help of the real Hubble radius. Instead, an effective ``Hubble radius" was introduced due to a kinetic coupling as analyzed in previous subsection. The primordial power spectrum of the field fluctuation $\delta\chi$ is expressed as
\begin{eqnarray}\label{P_deltachi}
 P_{\delta\chi} = \frac{k^3}{2\pi^2}|\frac{v_k}{z}|^2~.
\end{eqnarray}
Substituting the expressions of ${\cal Y}$ \eqref{Y_sol} into \eqref{P_deltachi} can give rise to the amplitude of the power spectra of interest.

\subsubsection{Case I: ${\cal Y} = (-{\cal H}_{\rm E}\tau)^{-(2+2 \epsilon)}$}

We first consider the first case: ${\cal Y} = (-{\cal H}_{\rm E}\tau)^{-(2+2 \epsilon)}$. In this case the friction term brought by the coefficient ${\cal Y}$ is similar to that of a ``de-Sitter expansion". From (\ref{sol}) it follows that on scales larger than the effective ``Hubble radius", the amplitude of the spectrum of $\delta\chi$ in our model is given by
\begin{eqnarray}\label{dchi1}
 \delta\chi \, = \, P_{\delta\chi}^{\frac{1}{2}} \, \simeq \, \frac{a_{\rm E}{H}_{\rm E}}{2\pi a}~,
\end{eqnarray}
which is almost constant in the emergent universe phase. Then we take into account the corrections of the $m^2$ term in the perturbation equation \eqref{eom_vk}. In particular, we choose
\begin{eqnarray}\label{F_caseI}
 {\cal F} = c_{\cal F} {\cal Y}^2~,
\end{eqnarray}
as an example. From the perturbation equation, we are able to calculate the spectral tilt $n_{\chi}$ of the primordial perturbations
\begin{eqnarray}\label{n chi}
n_{\chi}-1 \, \equiv \, \frac{d\ln P_{\delta\chi}}{d\ln k} \, = \, -2\epsilon +\frac{2c_{\cal F} a^2m^2}{3a_{\rm E}^2H_{\rm E}^2}~
\end{eqnarray}
where $m$ is the real mass of the $\chi$ field induced by the potential. From this result, we find that in order to ensure that the curvaton perturbations are nearly scale-invariant, one has to require $m \ll H_{\rm E}$. This is in agreement with the assumption we made in deriving the analytical solution to the perturbation equation.

\subsubsection{Case II: ${\cal Y} = (-{\cal H}_{\rm E}\tau)^{4+2\epsilon}$}

Then we consider the second case: ${\cal Y} = {(-{\cal H}_{\rm E}\tau})^{4+2\epsilon}$. In this case the friction term brought by the coefficient ${\cal Y}$ is similar to that of a ``matter-contraction". It turns out that the amplitude of the curvaton spectrum is given by
\begin{eqnarray}\label{dchi2}
 \delta\chi \, = \, P_{\delta\chi}^{\frac{1}{2}} \, \simeq \, \frac{1}{2\pi a_{\rm E} {\cal H}_{\rm E}^2|\tau|^3}~,
\end{eqnarray}
which is growing during the epoch of emergent universe. The amplitude of the curvaton perturbation stops increasing at the moment $\tau_{\rm E}$ which satisfies $\tau_{\rm E} = 2/{\cal H}_{\rm E}$ so that the universe can connect the dust-like expanding phase smoothly. Eventually we have $\delta\chi \simeq H_{\rm E}/16\pi$ at the super-Hubble scales after the emergent universe phase.

In this case, we would like to take the form of the coupling ${\cal F}$ as
\begin{eqnarray}\label{F_caseII}
 {\cal F} = c_{\cal F} {\cal Y}^{\frac{1}{2}} ~,
\end{eqnarray}
and then can obtain the same expression for the spectral index as Eq. \eqref{n chi}.

\subsection{The stability issue and constraints on background initial conditions}

Note that, it is important to examine the backreaction of the curvaton field upon the background dynamics, since we do not expect the curvaton to spoil the background emergent universe solution. Thus, we study the evolutions of the curvaton field with the kinetic couplings obtained above.

\subsubsection{Case I: ${\cal Y} = (-{\cal H}_{\rm E}\tau)^{-(2+2 \epsilon)}$}

To transform back to the frame of cosmic time, during the phase of emergent universe we have
\begin{eqnarray}\label{Y_caseI}
 {\cal Y} \simeq \left[ \frac{a_{\rm E}}{a_{\rm i}}
 H_{\rm E} (\tilde{t}_{\rm E}-t) \right]^{-(2+2\epsilon)}~,
\end{eqnarray}
with an integration constant $\tilde{t}_{\rm E} = t_{\rm E} +\frac{2a_{\rm i}}{a_{\rm E}H_{\rm E}}$ being introduced. Making use of \eqref{bar_psi_psi}, \eqref{scalefactor} and \eqref{Y_caseI}, we can reconstruct the form of the kinetic coupling ${\cal Y}$ as:
\begin{align}
& {\cal Y}(\bar\psi\psi) \simeq \nonumber\\
& \bigg\{ \frac{a_{\rm E}}{a_{\rm i}} H_{\rm E} \left[ \tilde{t}_{\rm E} -\frac{\ln\left( (1 +\frac{2\alpha{\cal C}}{3}) (\frac{\cal N}{a_{\rm E}^3\bar\psi\psi})^{\frac{\alpha{\cal C}}{3}} -1 \right)}{\alpha M_p}  \right] \bigg\}^{-(2+2\epsilon)} ~,
\end{align}
and in this approximate expression we take ${\cal Y}$ to approach unity when $\bar\psi\psi\rightarrow 0$. In this case, the curvaton field can recover the canonical form in the infrared regime.

Substituting the expression \eqref{Y_caseI} and \eqref{F_caseI} into the background equation of motion for the curvaton \eqref{eom_chi}, one can obtain two approximate solutions to the curvaton, which are
\begin{eqnarray}\label{chi_caseI}
 \chi(t) \sim (\tilde{t}_{\rm E}-t)^{\frac{c_{\cal F}a_{\rm i}^2m^2}{3a_{\rm E}^2H_{\rm E}^2}} ~,
 ~~{\rm or} ~~ \chi(t) \sim (\tilde{t}_{\rm E}-t)^3~,
\end{eqnarray}
respectively.

In Eq. \eqref{chi_caseI}, the first solution of $\chi$ determines a slow-evolving curvaton field and it implies that, during the emergent universe phase the curvaton field feels itself in a ``de-Sitter" background due to the kinetic coupling. In this regard, one can choose the curvaton field to be around the vacuum state as the initial condition for the background evolution. This choice is accompanied with the initial condition of quantum vacuum fluctuations. By inserting Eqs. \eqref{Y_caseI}, \eqref{F_caseI} and the first solution of \eqref{chi_caseI} into Eq. {densitypressure}, the corresponding energy density can be estimated as $\rho_\chi \sim (\tilde{t}_{\rm E}-t)^{-4}$. As a result, we can find that the contribution of the curvaton field is subdominant if we take it on the vacuum state initially. Along with the background evolution, the energy density of the curvaton field would grow slightly faster than the background density of the spinor field. Depending on the parameter choice of the curvaton field, the universe would either exit the emergent universe phase earlier than $t_{\rm E}$ due to the domination of the curvaton, or enter the post-emergent universe phase after $t_{\rm E}$ with both the spinor and the curvaton field evolving in paralleled trajectories. In both situations one can transform the curvaton fluctuation into the curvature perturbation through the curvaton mechanism as will be discussed in next section.

Then we consider the second solution of $\chi$ in Eq. \eqref{chi_caseI}. This solution implies that the curvaton field evolves rapidly and the initial value lies at large-valued regime. Its behavior is in analogy with the ``over-shoot" issue of inflationary cosmology that a scalar field may not feel the effect of the potential in a ``de-Sitter" background since initially its speed is chosen to be over large. In this case, the energy density of the curvaton becomes $\rho_\chi \sim (\tilde{t}_{\rm E}-t)^2$. Obviously, in this situation the curvaton field would have dominated over the background universe at very early times and thus spoils the scenario of emergent universe. We consider this background initial condition is not allowed in our model.

\subsubsection{Case II: ${\cal Y} = (-{\cal H}_{\rm E}\tau)^{4+2\epsilon}$}

In Case II, the coupling ${\cal Y}$ in the emergent universe phase can be approximately expressed as
\begin{eqnarray}\label{Y_caseII}
 {\cal Y} \simeq \left[ \frac{a_{\rm E}}{a_{\rm i}}
 H_{\rm E} (\tilde{t}_{\rm E}-t) \right]^{4+2\epsilon}~.
\end{eqnarray}
Again, to combine Eqs. \eqref{bar_psi_psi}, \eqref{scalefactor} and \eqref{Y_caseI}, the form of the kinetic coupling ${\cal Y}$ can be reconstructed as follows,
\begin{align}
& {\cal Y}(\bar\psi\psi) \simeq \nonumber\\
& \bigg\{ \frac{a_{\rm E}}{a_{\rm i}} H_{\rm E} \left[ \tilde{t}_{\rm E} -\frac{\ln\left( (1 +\frac{2\alpha{\cal C}}{3}) (\frac{\cal N}{a_{\rm E}^3\bar\psi\psi})^{\frac{\alpha{\cal C}}{3}} -1 \right)}{\alpha M_p}  \right] \bigg\}^{4+2\epsilon}~,
\end{align}
which can recover the canonical kinetic term for the curvaton $\chi$ in the low energy limit as well.

We then solve the equation of motion for the curvaton \eqref{eom_chi} and obtain two approximate solutions as follows,
\begin{eqnarray}\label{chi_caseII}
 \chi(t) \sim (\tilde{t}_{\rm E}-t)^{-\frac{c_{\cal F}a_{\rm i}^2m^2}{3a_{\rm E}^2H_{\rm E}^2}} ~,
 ~~{\rm or} ~~ \chi(t) \sim (\tilde{t}_{\rm E}-t)^{-3}~,
\end{eqnarray}
respectively.

In this case, the first solution of \eqref{chi_caseII} gives rise to the energy density of the curvaton field as $\rho_\chi \sim (\tilde{t}_{\rm E}-t)^2$. The second solution of \eqref{chi_caseII} leads to $\rho_\chi \sim (\tilde{t}_{\rm E}-t)^{-4}$. Note that, both two solutions imply that the curvaton field lies on the vacuum state at initial moment since the power indices are negatively definite. The difference is that for the first solution the curvaton field evolves slowly and the corresponding energy density would spend much more time to catch up that of the background spinor field.

As a side remark, we would like to point out that a nearly scale-invariant power spectrum of the curvaton can be obtained merely by the ${\cal F}$ parameter as well. For simplicity, we can choose the canonical kinetic term with ${\cal Y}= 1$ and take the quadratic potential $V(\chi)=\frac{1}{2}m^2\chi^2$. Notice that at the stage of emergent universe, the scale factor is almost constant, following Eq. \eqref{eom_vk}, the scale invariance requires that ${\cal F}(\tau)\sim -\frac{2}{a^{2}_{\rm E}m^2\tau^2}$. As the sign of this term is negative, the scenario suffers from a severe tachyonic instability. We will not address this case in the paper.

\section{Numerical Computation}\label{Sec:numerics}

\begin{figure}
\includegraphics[scale=0.35]{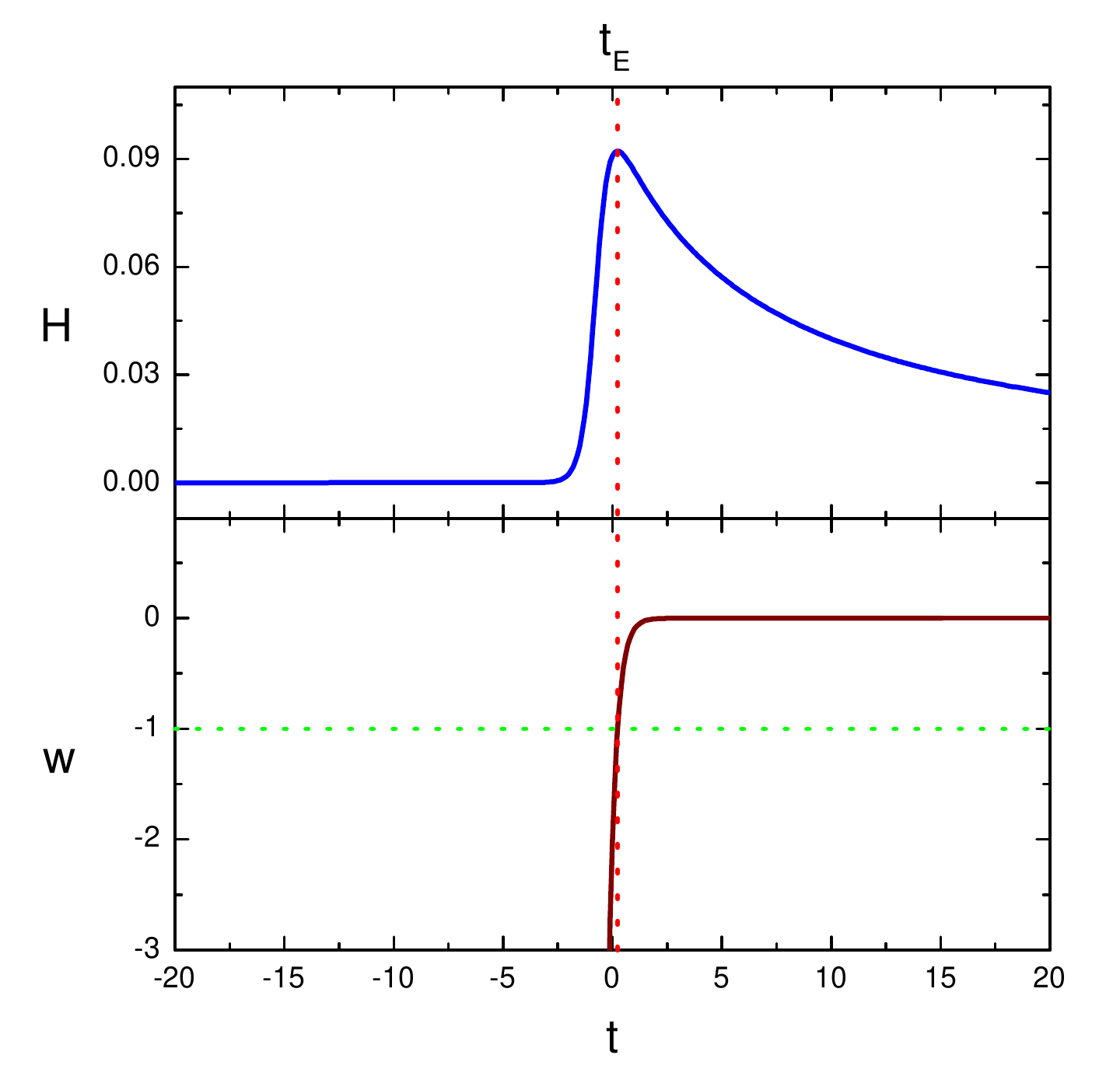}
\caption{Numerical plot of the evolutions of the Hubble parameter $H$ and the background EoS $w$ as a function of cosmic time in the model under consideration. In the numerical calculation, we take $\alpha=3$, ${\cal C}=10$. All dimensional parameters are of Planck units. }
\label{Fig:Hw}
\end{figure}

In this section, we numerically examine the dynamics of the model under consideration. We show the evolutions of the Hubble parameter $H$ and the EoS $w$ of the background universe in Fig. \ref{Fig:Hw}. The specific parameterizations follow Eq. \eqref{Hubble_EU} with the model parameters being taken as: $\alpha=3$, ${\cal C}=10$ and ${\cal N}=0.1$. We use the blue solid line to depict the Hubble parameter and the wine solid line to represent for the EoS. At the moment of $t_{\rm E}$, the Hubble parameter reaches the maximal value. The EoS $w$ initially is much less than $-1$, and evolve through the cosmological constant boundary (denoted by the light green dotted line in the lower panel) at the moment $t_{\rm E}$, and then approaches zero afterwards.

\begin{figure}
\includegraphics[scale=0.35]{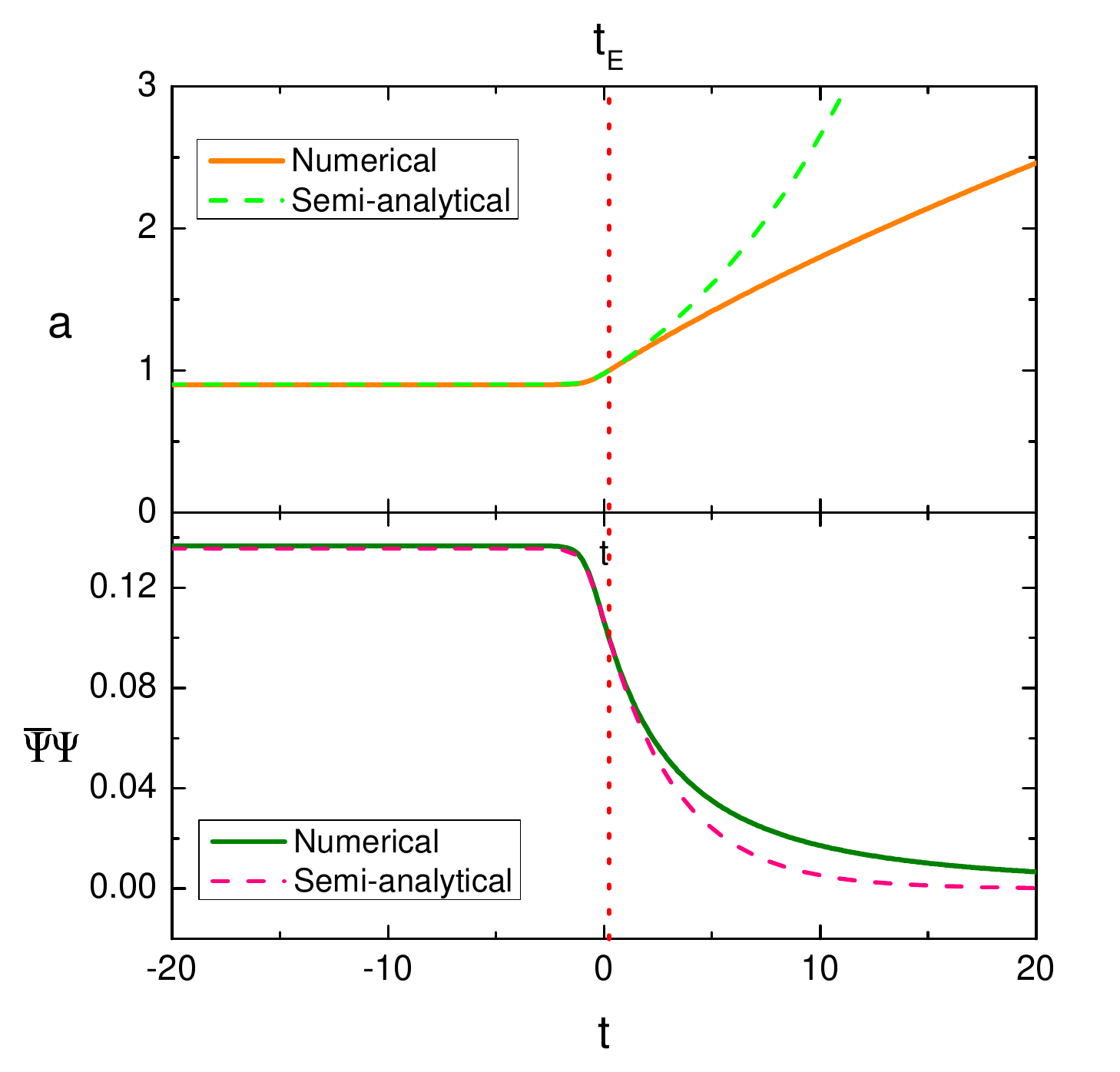}
\caption{Comparison of the numerical result and the semi-analytical estimate of the evolutions of the scale factor $a$ and the scalar bilinear $\bar\psi\psi$ as a function of cosmic time in the model under consideration. In the numerical calculation, the values of $\alpha$ and ${\cal C}$ are the same as provided in Fig. \ref{Fig:Hw} and ${\cal N}=0.1$ for the scalar bilinear $\bar\psi\psi$. All dimensional parameters are of Planck units. }
\label{Fig:app}
\end{figure}

Having known the dynamics of the Hubble parameter, one can integrate out the evolutions of the scale factor and then the scalar bilinear $\bar\psi\psi$ exactly. In order to better understand the qualitative calculation performed in previous sections, we compare the numerical result and the semi-analytical estimate of the evolutions of the scale factor $a$ and the scalar bilinear $\bar\psi\psi$ in Fig. \ref{Fig:app}. From the figure, one can explicitly find that the scale factor $a$ approaches to a non-vanishing minimal value in the far past, and connects the matter-like expansion when $t$ is larger than $t_{\rm E}$. The numerical result (the orange solid line) and the semi-analytical estimate (the light green dashed line) in the upper panel coincide before $t_{\rm E}$. This feature nicely shows that the previous estimate is in good agreement with the realistic situation in the phase of emergent universe. These two curves deviate from each other in the matter-like expanding phase.

\begin{figure}
\includegraphics[scale=0.35]{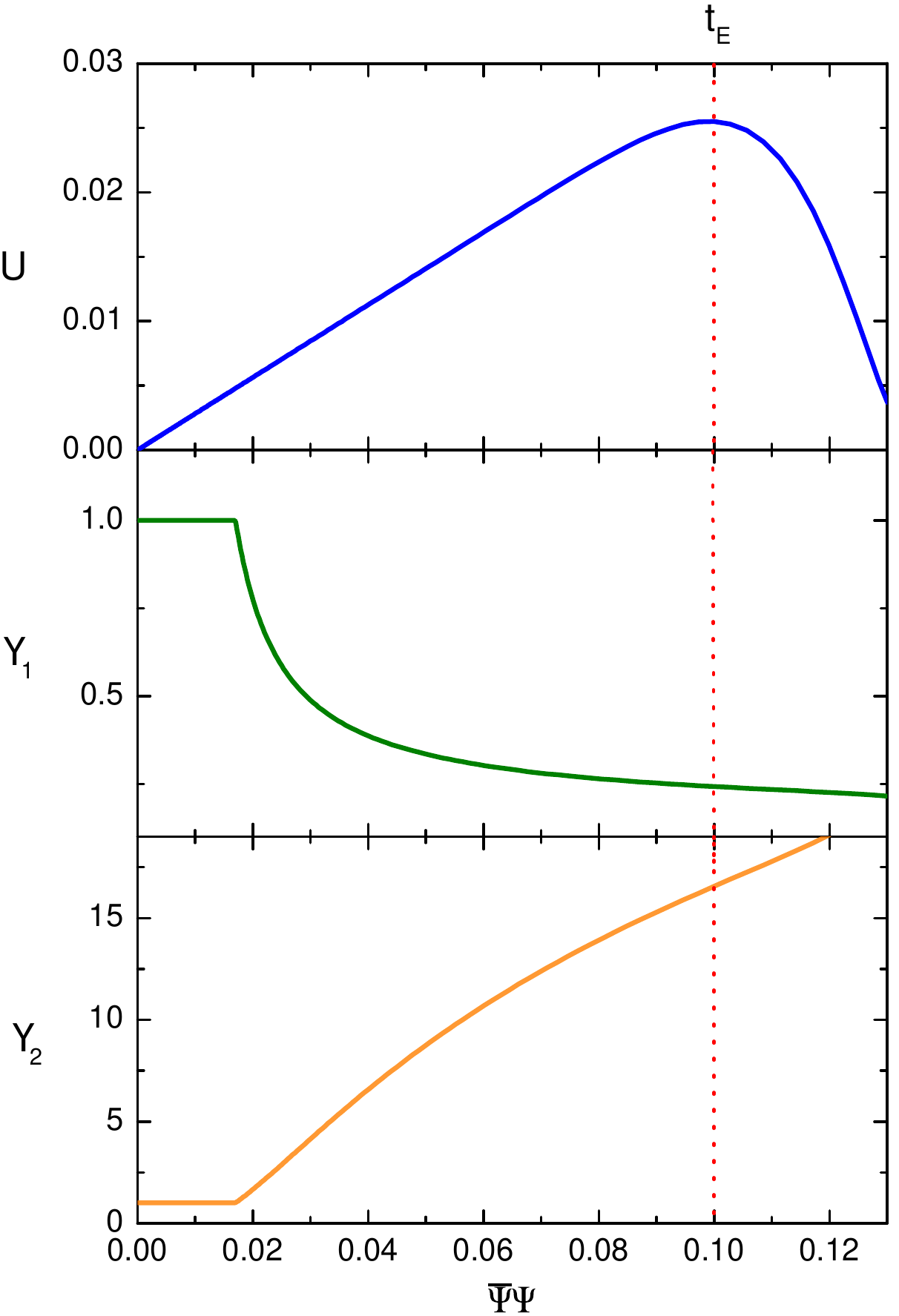}
\caption{Numerical plot of the potential $U$ and the kinetic coupling ${\cal Y}$ as a function of the bilinear $\bar\psi\psi$ in the model under consideration. We use ${\cal Y}_1$ to denote the solution in Case I and ${\cal Y}_2$ to denote the solution in Case II. In the numerical calculation, we take the values of $\alpha$, ${\cal C}$ and ${\cal N}$ the same as in Fig. \ref{Fig:Hw}. All dimensional parameters are of Planck units. }
\label{Fig:UYY}
\end{figure}

Afterwards, we can compute the potential $U$ as well as the kinetic coupling coefficient ${\cal Y}$ as a function of $\bar\psi\psi$. The corresponding result is presented in Fig. \ref{Fig:UYY}. Since the condition of scale invariance for curvaton fluctuations yields two possible solutions to the kinetic coupling ${\cal Y}$, we denote ${\cal Y}_1$ (the green solid line the middle panel) as the solution in Case I and ${\cal Y}_2$ (the orange solid line in the lower panel) as the solution in Case II. Additionally, when the universe exits the quasi-Minkowski phase, we expect that the curvaton field can decouple from the cosmic spinor. Therefore, we take a cutoff in the numerical estimate by choosing the convention ${\cal Y}=1$ in the regular expanding phase. One can read from Fig. \ref{Fig:UYY} that in Case I the kinetic coupling ${\cal Y}$ decreases along the bilinear $\bar\psi\psi$, but in Case II ${\cal Y}$ becomes large in the far past. Due to different dynamics of ${\cal Y}$, the effective ``Hubble radius" shows different features in the emergent universe phase.

\section{Curvature perturbation via curvaton mechanism}\label{Sec:curvaton}

To proceed, we calculate the curvature perturbations generated by the curvaton field through the curvaton mechanism \cite{Lyth:2001nq, Linde:1996gt}. After the emergent universe phase has ceased, we assume that the cosmic spinor field decays to radiation according to the curvaton mechanism. During this period, the energy density of the universe is composed of the radiation energy density $\rho_r$ and the energy density of the curvaton field $\rho_\chi$. During this epoch, the curvature perturbation is seeded by the iso-curvature fluctuations since the pressure perturbations are non-adiabatic. This process ends when the perturbations become adiabatic again, which corresponds to either the epoch of curvaton domination, or that of curvaton decay. The final curvature perturbations can be calculated at the moment that the Hubble rate is comparable with the decay rate based on the assumption of perturbatively instantaneous reheating of the curvaton field.

We can simply consider the component curvature perturbations $\zeta_\chi$ and $\zeta_\psi$ on the slices of uniform curvaton density and radiation density separately. However, following Ref. \cite{Cai:2012yf}, the perturbation contributed by the spinor field leads to a blue spectrum and thus is subdominant at large scales of observable interest. We can neglect this part of contribution and thus the curvature perturbation on super-Hubble scales takes the form as follows,
\begin{eqnarray}\label{curvaturetotal}
 \zeta= \tilde{r} \zeta_{\chi}~,
\end{eqnarray}
with a generic transfer efficiency parameter being defined by \cite{Cai:2010rt},
\begin{eqnarray}
 \tilde{r} \equiv \frac{3(1+w_{\chi})\rho_{\chi}}{4\rho_{r}+3(1+w_{\chi})\rho_{\chi}}~,
\end{eqnarray}
where $w_{\chi}\equiv P_{\chi}/\rho_{\chi}$ is defined as the EoS of the curvaton field\footnote{The curvaton mechanism with a generalized EoS is motivated by some nonstandard curvaton models, such as the K-essence version \cite{Cai:2009hw}, the DBI form \cite{Li:2008fma}, or a polynomial potential \cite{Huang:2008bg}.}.

On super-Hubble scales, the component curvature perturbation on uniform density slice can be written as
\begin{equation}\label{deltaN}
  \zeta_\chi(x)=\delta N(x)+\frac{1}{3}\int_{\bar\rho_\chi(t)}^{\rho_\chi(x)}\frac{d\tilde\rho_\chi}{\tilde\rho_\chi+P_\chi(\tilde\rho_\chi)}~.
\end{equation}
As a first step, one need to write down the relation between the curvaton fluctuation $\delta\chi$ and the corresponding curvature perturbation. By taking the spatially flat slice, if we want to consider a general EoS for the curvaton, Eq. \eqref{deltaN} becomes
\begin{equation}\label{deltaNchi}
  \rho_\chi=\bar\rho_\chi e^{3(1+w_\chi)\zeta_\chi}~,
\end{equation}
in the neighborhood of the curvaton decay hyper-surface.

Consider the curvaton perturbation that is initially originated from vacuum fluctuations inside the Hubble horizon. These
perturbation modes could satisfy a Gaussian distribution at the Hubble exit. In this case, we have
\begin{equation}
  \chi_*=\bar\chi_*+\delta\chi_*~.
\end{equation}
This Hubble-crossing value can be related to the initial amplitude of curvaton oscillation, which takes the form \cite{Lyth:2002my}:
\begin{equation}
  g(\chi_*)=g(\bar\chi_*+\delta\chi_*)=\bar{g} +\sum_{n=1}^{\infty}\frac{g^{(n)}}{n!}\left(\frac{\delta\chi}{g_{,\chi}}\right)^n ~.
\end{equation}
The detailed form of $g(\chi_*)$ is a model-dependent function which is determined by the explicit potential of the curvaton
field in the non-relativistic limit. For example, if the curvaton potential is quadratic as given by Eq. \eqref{V_chi} until the curvaton decay, then we have $g(\chi_*)\propto \chi_*$. Consequently, the energy density can be approximately expressed as
\begin{equation}\label{curvatondensity}
  \rho_\chi = \frac{1}{2} {\cal F} m^2 ( \bar{g}^2 + 2\bar{g} \delta\chi ) + O(\delta\chi^2)~,
\end{equation}
up to linear order. Therefore, the combination of Eqs. \eqref{deltaNchi} and \eqref{curvatondensity} leads to the component curvature perturbation as
\begin{eqnarray}\label{zeta_chi}
  \zeta_{\chi}=\frac{2}{3(1+w_\chi)}\frac{\delta\chi}{\bar\chi}~.
\end{eqnarray}

The second step of the calculation is to relate $\zeta_\chi$ to $\zeta$. In the sudden decay approximation, the relation is quite simple and can be computed analytically. We assume the curvaton decays on a uniform total density hyper-surface $H=\Gamma$, where $\Gamma$ is the decay rate of the curvaton. This can be realized in the framework of particle physics, namely the curvaton has a ``life time'' and then decays to particles through their coupling terms. Then on the curvaton
decay hyper-surface we have $\rho_r+\rho_\chi=\bar\rho$. In a generic case the EoS of the curvaton depends on the detailed evolution of the curvaton. Namely, an oscillating curvaton with a potential of form $m^2\chi^2$, it behaves as a pressureless field fluid. Here, we make use of Eqs. \eqref{curvaturetotal} and \eqref{zeta_chi}, and then get
\begin{eqnarray}\label{zeta_total}
 \zeta = \frac{2\tilde{r}}{3(1+w_\chi)}\frac{\delta\chi}{\bar\chi}~,
\end{eqnarray}
Note that, the generic transfer efficiency parameter can be further simplified as,
\begin{eqnarray}
 \tilde{r} = \frac{3(1+w_\chi)\Omega_\chi}{4+(-1+3w_\chi)\Omega_\chi}~,
\end{eqnarray}
where $\Omega_{\chi}=\bar\rho_{\chi}/(\bar\rho_r+\bar\rho_{\chi})$ is the dimensionless density parameter for the curvaton at the decay moment.

Eventually, we can calculate the power spectrum of primordial curvature perturbation seeded by the curvaton fluctuations in the model of spinor emergent universe cosmology. As we have two possible solutions to generate the curvaton fluctuations as was analyzed in previous section. In the following we discuss both two cases.

For Case I, we can choose initially the curvaton sits on the vacuum state and slowly evolves during the emergent universe phase. The energy density of the curvaton grows as $\rho_\chi \sim (\tilde{t}_{\rm E}-t)^{-4}$ with the cosmic time $t$ approaches $t_{\rm E}$ from negative infinity. Therefore, the detailed value of $\tilde{r}$ depends on whether the universe exits the epoch of emergent universe due to the domination of $\rho_\chi$ or the dynamics of the cosmic spinor $\psi$ automatically. By combining Eqs. \eqref{dchi1} and \eqref{zeta_total}, the power spectrum of primordial curvature perturbation is given by
\begin{eqnarray}
 P_\zeta = \zeta^2 = \frac{\tilde{r}^2 H_{\rm E}^2}{9\pi^2(1+w_\chi)^2\bar\chi^2}~,
\end{eqnarray}
with the spectral index inheriting from Eq. \eqref{n chi}.

Similarly, for Case II, the power spectrum of primordial curvature perturbation can be calculated by inserting the solution \eqref{dchi2} into \eqref{zeta_total}
\begin{eqnarray}
 P_\zeta = \frac{\tilde{r}^2H_{\rm E}^2}{576\pi^2(1+w_\chi)^2\bar\chi^2}~,
\end{eqnarray}
of which the amplitude is relatively lower than that obtained in the previous case.

We would like to emphasize two key features of the above calculation. First, the curvature perturbation $\zeta$ and $\zeta_\chi$ are gauge invariant. Therefore, although we study their dynamics in the gauge of spatially flat slice, their values are irrelevant to the gauge choice. Second, in the approximation of sudden decay, $\zeta$ is conserved right after the universe evolves through the curvaton decay hyper-surface and the curvature perturbation should be calculated exactly on this hyper-surface. This is because, if one calculates at any earlier time, $\zeta$ is not conserved due to the contribution of iso-curvature perturbation; if one calculates at any later time, the information stored in the curvaton-induced curvature perturbation $\zeta_\chi$ would have been lost.

\section{Conclusions}\label{Sec:conclusions}

To conclude, in the present paper we have studied the possible causal mechanisms of generating scale-invariant primordial power spectrum in the cosmology of spinor emergent universe. Specifically, we introduce another light scalar field which kinetically couples to the cosmic spinor field and use the scalar field to generate iso-curvature perturbation in the primordial era. Because of the kinetic coupling terms, the field fluctuations of the curvaton feel themselves to evolve within a ``de-Sitter"-like background or a ``matter-contraction" one. Therefore, there exists an effective ``Hubble radius" for the iso-curvature modes to become classical perturbation during the emergent universe phase. After having propagated for enough long time on the ``super-Hubble" scales, these iso-curvature modes can be converted into curvature perturbation at the moment of curvaton decay.

The cosmology of emergent universe, as an alternative to inflationary cosmology, can avoid the initial spacetime singularity. Its implement may be inspired by string theory such as the string gas cosmology, but the phenomenological model-building often suffers from detailed difficulties, namely the instability of the backreaction, or the graceful-exit issue. The model considered in the present work has provided a representative example to illustrate that the emergent universe cosmology could be free of the instabilities and the issue of graceful-exit. Based on this model, there are many interesting topics which deserve a further investigation in the future. For example, one would expect to constrain the parameter space of this model by confronting with the latest cosmological observations; it is expected to relate the model of spinor emergent universe to the spinor formalism in fundamental theories such as the open string field theory; the dynamics of nonlinear perturbations in this model and the related stability issue ought to be analyzed. We would like to address these interesting topics in the future study.

\begin{acknowledgments}
We thank Robert Brandenberger, Hong Li and Jun-Qing Xia for useful discussions. The work of CYF is supported in part by Department of Physics in McGill University. The research of YW and XZ is supported in part by the National Science Foundation of China under Grants No. 11121092, 11033005 and 11375202.
\end{acknowledgments}

\end{document}